\documentclass[10pt,aps,prd,superscriptaddress,nofootinbib,nobibnotes,longbibliography,floatfix,twocolumn]{revtex4-2}

\usepackage{bm}
\usepackage{mathtools,
amsmath,
amssymb,
amsfonts,
mathrsfs,
chngcntr,
multirow}

\let\cc\corresponds
\let\corresponds\relax
\usepackage{mathabx}
\let\corresponds\cc

\usepackage[utf8]{inputenc}
\usepackage[T1]{fontenc}
\usepackage[dvipsnames]{xcolor}
\usepackage[unicode]{hyperref}
\hypersetup{colorlinks=true, citecolor=MidnightBlue,
            linkcolor=MidnightBlue, urlcolor=MidnightBlue, linktocpage=true}
\usepackage[normalem]{ulem}
\usepackage{pifont} 
\usepackage[inline]{enumitem}
\usepackage{soul} 
\usepackage{derivative}
\usepackage{empheq}
\usepackage{tcolorbox}

\usepackage{orcidlink}

\newcommand{\orcid}[1]{\href{https://orcid.org/#1}{\includegraphics[width=10pt]{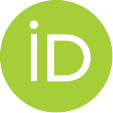}}}

\newlist{inlinelist}{enumerate}{1}
\setlist[inlinelist,1]{label=\,(\arabic*), before=\unskip, after=\unskip, itemsep=0pt, left=0pt}

\begin{document}

\title{Ringdown spectroscopy of phenomenologically modified black holes}

\author{Spyros Thomopoulos\,\orcid{0009-0007-5191-3693}}
\email{spyros.thomop@gmail.com}
\affiliation{Physics Department, National Technical University of Athens, Zografou, Greece}

\author{Sebastian H. V\"olkel\,\orcid{0000-0002-9432-7690}}
\email{sebastian.voelkel@aei.mpg.de}
\affiliation{Max Planck Institute for Gravitational Physics (Albert Einstein Institute), Am M\"uhlenberg 1, Potsdam 14476, Germany}

\author{Harald P. Pfeiffer\,\orcid{0000-0001-9288-519X}}
\affiliation{Max Planck Institute for Gravitational Physics (Albert Einstein Institute), Am M\"uhlenberg 1, Potsdam 14476, Germany}

\date{\today}

\begin{abstract}
The characteristic oscillations of black holes, as described by their quasinormal mode (QNM) spectrum, play a fundamental role in testing general relativity with gravitational waves. 
The so-called parametrized QNM framework was introduced to predict perturbative changes in the spectrum due to small deviations from general relativity. 
In this work, we extend the framework to model time domain signals and study the excitation of quasinormal modes from the time evolution of initial data. 
Specifically, we quantify whether the perturbative eigenvalue predictions agree with extracting excited quasinormal modes from such simulations. 
Addressing this issue is particularly important in the context of agnostic ringdown tests and the possible presence of spectral instabilities, which may diminish the promises of black hole spectroscopy. 
We find that the extracted quasinormal modes agree well with the perturbative predictions, underlining that these types of modifications can, in principle, be robustly tested from real observations. 
Moreover, we also report the importance of including late-time tails for accurate mode extractions. 
Finally, we provide a WKB-inspired analysis supporting the importance of the peak of the scattering potential and show robust scaling relations. 

\end{abstract}

\maketitle

\section{Introduction}\label{intro}

Einstein's theory of general relativity (GR) is one of the best-tested theories in modern physics~\cite{Einstein:1916vd}. 
Many important predictions have been confirmed by observations, starting from Solar System tests like the perihelion precession of Mercury, the gravitational deflection of light, and more~\cite{Will:2014kxa}, up to strong-gravity dynamics, like the Hulse-Taylor binary confirming indirectly the existence of gravitational waves~\cite{Hulse:1974eb}. 
In more recent history, black holes have emerged in a central role in large-scale endeavors, such as the direct detection of gravitational waves from compact binary mergers by the LIGO-Virgo-KAGRA Collaboration~\cite{LIGOScientific:2016aoc,LIGOScientific:2017vwq} or the direct imaging of supermassive black holes by the Event Horizon Telescope Collaboration~\cite{EventHorizonTelescope:2019dse,EventHorizonTelescope:2022wkp}. 

Despite its success, increasing tensions in cosmological parameters could, in principle, also point towards possible deviations from GR on large scales~\cite{Kamionkowski:2022pkx}. 
Moreover, the unclear resolution of singularities in black holes and the open quest for quantum gravity continue to challenge us to rethink and test gravity.  
The revolutionary access to gravitational waves emitted from merging black holes probes the strong and dynamical aspect of GR. 
It can, in principle, be modeled with arbitrary precision, which is exceptional for astrophysical systems. 

The intriguing possibility to study black holes by analyzing their gravitational wave oscillation spectrum is commonly known as black hole spectroscopy~\cite{Dreyer:2003bv}; see Refs.~\cite{Kokkotas:1999bd,Nollert:1999ji,Berti:2009kk,Konoplya:2011qq,Franchini:2023eda,Berti:2025hly} for reviews on the topic. 
This spectrum of quasinormal modes (QNMs) contains infinitely many characteristic frequencies and damping times labeled by angular numbers $\ell,m$ and overtones $n$. 
If GR is the correct theory of gravity, and if the assumptions of no-hair theorems are valid~\cite{Israel:1967wq,Hawking:1971vc,Carter:1971zc,Robinson:1975bv}, all QNMs of an isolated black hole are uniquely determined by its mass and spin. 
Therefore, measuring more than one QNM allows one to test our fundamental understanding of black holes and GR itself. 

While pioneering works in black hole perturbation theory span more than half a century~\cite{Regge:1957td,Zerilli:1970se, Teukolsky:1973ha}, the challenging task of extracting them from high-accuracy time domain simulations has received a lot of attention in recent years; see Refs.~\cite{Giesler:2019uxc,Sberna:2021eui,Cheung:2022rbm,Mitman:2022qdl,Konoplya:2022pbc,Baibhav:2023clw,Nee:2023osy,Cheung:2023vki,Zhu:2023mzv,Redondo-Yuste:2023ipg,Silva:2024ffz,Giesler:2024hcr,Mitman:2025hgy} for some examples. 
One of the key challenges in black hole spectroscopy, especially for binary mergers, is that the QNMs describe a strongly damped system, making it very challenging to use ordinary mode extraction techniques. 
Moreover, QNMs are only a good description of a perturbed black hole at intermediate times. 
When excited through binary black hole mergers, the early part of the signal includes nonlinear imprints of the full Einstein field equations and a prompt response that depends on the details of the initial perturbations. 
At very late times, a power-law tail overtakes the exponential decay of any QNM~\cite{Price:1971fb,Leaver:1986gd,Gundlach:1993tp,Barack:1998bw}. 
Another complication arises from the fact that QNMs do not form a complete set. 
Thus, not even the linearized problem can be easily addressed. 

In recent years, several works revisited the robustness of the QNM spectrum with respect to small changes in the effective potential appearing in the relevant equations, which were first studied in Refs.~\cite{Nollert:1996rf,Nollert:1998ys}, reexamined in Ref.~\cite{Daghigh_2020}, and investigated in the context of environmental effects in Ref.~\cite{Barausse:2014tra}. 
In a series of works, it has been shown that some types of modifications of the standard GR problem can yield very different QNMs, especially for overtones~\cite{Jaramillo:2020tuu,Jaramillo:2021tmt,Cheung:2021bol,Sarkar:2023rhp,Courty:2023rxk}. 
However, what modifications can be considered small in terms of the physical system is nontrivial; see Refs.~\cite{Konoplya:2022pbc,Cardoso:2024mrw} for different aspects. 
This raises some concerns for black hole spectroscopy, where one relies on establishing a connection between the QNMs from a potentially unstable eigenvalue problem and their extraction from evolving perturbations in the time domain; see Refs.~\cite{Nollert:1996rf,Nollert:1998ys} for seminal studies. 
One theory-agnostic approach that introduces small phenomenological modifications to the perturbation equations is the so-called parametrized QNM framework~\cite{Cardoso:2019mqo, McManus:2019ulj,Kimura:2020mrh,Volkel:2022aca,Franchini:2022axs,Hirano:2024fgp}, 
which provides a perturbative correction to the QNMs as a function of modification parameters. 
While the validity of the perturbative predictions has been well tested, no related study has been conducted on extracting the perturbed QNMs from the time domain.  

We quantify to what extent the modifications introduced in the parametrized QNM framework can also be robustly inferred in the time domain. 
We first introduce a time dependency to the perturbation equations to address this objective and then evolve initial data to excite QNMs.  
To quantify the QNM content of the signals, we use a superposition of damped sinusoids where each mode has four independent parameters. 
Furthermore, we show that the stability of the QNM extraction can be improved by including the Price power-law tail~\cite{Price:1971fb,Leaver:1986gd,Gundlach:1993tp,Barack:1998bw}.

Our results demonstrate that the fundamental mode is in excellent agreement with the perturbative predictions from the parametrized QNM framework. 
This holds for various possible modifications, both for one parameter at a time and multiple parameters at a time.  
Our study is, therefore, also an independent verification of the framework's predictions and its most systematic analysis in the time domain. 
More importantly, it shows that it is not impacted, at least not in practice, by possible problems related to spectral stability. 
As a complementary aspect, we also look for correlations between the extracted QNM parameters and the local properties of the peak of the perturbation potential which is expected from Wentzel-Kramers-Brillouin (WKB) theory~\cite{Schutz:1985km,Iyer:1986np,Iyer:1986nq,Kokkotas:1988fm,Seidel:1989bp,Konoplya:2003ii,Matyjasek:2017psv}. 
We find a strong relation between the real part of the QNM fundamental mode and the height of the potential barrier, which is very insensitive to the specific modification and can thus be robustly inferred. 

The rest of this paper is organized as follows. 
Our methods are outlined in Sec.~\ref{methods}. 
Applications and results can be found in Sec.~\ref{applications}. 
We discuss our findings in Sec.~\ref{discussion}, and provide conclusions in Sec.~\ref{conclusions}. 
We use units in which $G=c=1$.

\section{Methods}\label{methods}

In this section, we present a brief summary of the parametrized QNM framework in Sec.~\ref{ss:pQNMf}, introduce its extension to the time domain
in Sec.~\ref{ss:time_evol}, and outline the ringdown fitting procedure in Sec.~\ref{ss:rd_modeling}.

\subsection{Parametrized QNM framework}\label{ss:pQNMf}

In this work, we adopt the parametrized QNM framework ~\cite{Cardoso:2019mqo, McManus:2019ulj,Kimura:2020mrh,Volkel:2022aca,Franchini:2022axs,Hirano:2024fgp} valid for metric perturbations around the Schwarzschild black hole. 
Its working hypothesis is that the imprint of a beyond-GR theory to the observable QNM spectrum is small and can be described by modifying the GR perturbation equations as follows. 
The framework introduces phenomenological deviations of the perturbation potential of metric perturbations and connects them to small shifts in the QNM spectrum. 
These modifications consist of adding various powers of $1/r$ to the perturbation potential, which is motivated by theory-specific examples. 
In its more general setup, it contains $N$ coupled fields, e.g., (scalar, vector, or tensor) $\bm{\Phi}=\left\{\Phi_i\right\}, i=1,\cdots,N$. 

As a representative case, we only consider a single-field case described by
\begin{align}\label{eq:def_PF}
f\odv{}{r}\left( f\odv{{\Phi}(r)}{r}  \right) + \left[ \omega^2 - f(r) {V(r)} \right]{\Phi}(r) = 0 \,,
\end{align}
where $f=1-r_\text{H}/r$, $r_\text{H}$ is the location of the event horizon, $\omega$ is the complex QNM frequency and $V(r)$ is the effective potential given by
\begin{align}
\label{eq:potential}
V(r) =& V_\mathrm{GR}(r) + \delta V(r), \\ 
\label{eq:potential_mod}
\delta V(r) =& \frac{1}{r_\text{H}^2} \sum_{k=0}^{\infty} \alpha^{(k)}\left(\frac{r_\text{H}}{r}\right)^k \,.
\end{align}
In this work, $V_\mathrm{GR}(r)$ is given by the Regge-Wheeler potential (axial metric perturbations) 
\begin{align}
V_\mathrm{GR}(r)  =& \frac{\ell(\ell+1)}{r^2} - \frac{3r_\text{H}}{r^3}\,,
\end{align}
and $\delta V(r)$ is the modification to the GR potential. 
Note the unusual convention of not including $f$ in the definition of the potential term.  

The modifications are described by the power $k$ and their corresponding amplitude $\alpha^{(k)}$, which is generally complex and can depend on $\omega$. 
The smallness of the modification can be captured by the condition~\cite{McManus:2019ulj} 
\begin{align}\label{eq:criterion}
\alpha^{(k)} \ll (1+1/k)^k(k+1)\,.
\end{align}
We note that the structures of polar perturbations and scalar test fields are qualitatively similar, so we do not consider them in this work. 

The solution of Eq.~\eqref{eq:def_PF} corresponding to the GR case can be described by QNMs $\omega_\mathrm{GR}$, which are the eigenvalues of the boundary value problem with outgoing waves at spatial infinity and ingoing waves at the horizon. 
Small deviations from the GR potential of this kind only lead to small deviations from the GR QNM. 
Thus, one can expand $\omega$ around $\omega_\mathrm{GR}$ up to quadratic order
\begin{align}\label{eq:omega_expansion}
\omega \approx \omega_\mathrm{GR} + \alpha^{(k)}d_{(k)} +\alpha^{(k)}\partial_\omega\alpha^{(s)} d_{(k)}d_{(s)} 
+ \frac{1}{2}\alpha^{(k)}\alpha^{(s)} e_{(ks)} \,,
\end{align}
where Einstein's convention is being used. 
We omit the label $\ell m n$ to avoid cluttering indices, which is assumed implicitly. 
The most important aspect of this equation is that the sets of coefficients $d_{(k)}$ and $e_{(ks)}$ are universal. 
They do not depend on the amplitude of the modifications but only on the unperturbed potential. 
Their numerical values depend on the perturbing field (scalar, axial, polar), but once provided, one can use them to easily study deviations with varying $\alpha$'s. 
The values up to quadratic order (ignoring the possible dependence on frequency here) have been calculated with the direct integration method in Refs.~\cite{Cardoso:2019mqo,McManus:2019ulj,Hirano:2024fgp}, as well as from the continued fraction method in Refs.~\cite{Volkel:2022aca,Cano:2024jkd}. 
We use the coefficients from the \texttt{GitHub}  repository~\cite{sebastian_volkel_2024_14001739} presented in Ref.~\cite{Cano:2024jkd}, which also extended the parametrized QNM framework beyond Teukolsky.

\subsection{Time evolution} \label{ss:time_evol}

The parametrized QNM framework of Sec.~\ref{ss:pQNMf} has been widely studied as an eigenvalue problem, which cannot represent late-time tails or initial data dependencies of the early ringdown stage.
To include such features, one should seek to represent the ringdown in the time domain.
Therefore, we need to find a way to extend the parametrized QNM framework, or at least a subset of possible scenarios, to the time domain. 
Since the framework starts already on the level of the eigenvalues in the perturbation equations, it cannot easily be derived from a more general context. 
Therefore, we make the following assumption valid for the theory-agnostic aspect of tackling the ringdown analysis in Sec.~\ref{applications}. 

We assume that the $\omega^2$ term of Eq.~\eqref{eq:def_PF} arises from a Fourier transformation of a second time derivative, which is true for GR. 
Moreover, we also assume that there are no other frequency-dependent terms. 
Those naturally arise for slowly rotating black holes, even within GR and could potentially appear if the background metric is not Schwarschild~\cite{Franchini:2022axs}.
Their presence would contaminate our time domain potential with additional terms containing time derivatives. 
With these assumptions, the wave equation we need to solve is given by
\begin{align}\label{eq:def_time_domain}
    \left[-\odv[order=2]{}{t} + \odv[order=2]{}{x}  - f(r) V(r) \right] \Phi(t, x) = 0 \,.
\end{align}
where the tortoise coordinate is defined as
\begin{align}
x = r + r_\mathrm{H} \ln\left(\frac{r}{r_\mathrm{H}} - 1 \right)\,.
\end{align}
GR modifications enter the potential $V(r)$ as defined in Eq.~\eqref{eq:potential}.

To solve Eq.~\eqref{eq:def_time_domain} for some initial data, we adopt a finite difference scheme central in time and space, as presented in Ref.~\cite{Nee:2023osy} 
\begin{align} 
\Phi^{j+1}_{i} = 2\Phi^{j}_{i}-\Phi^{j-1}_{i} + \frac{\Delta t^2}{\Delta x^{2}} \left(\Phi^{j}_{i+1} + \Phi^{j}_{i-1} - 2\Phi^{j}_{i}\right) \nonumber \\ 
 - \Delta t^2 \Phi^{j}_{i} \cdot V_{i},
\end{align}
where $\Phi^j_i=\Phi(t_j, x_i), V_{i} = V(x_i)$ and $\Delta t, \Delta x$ are the step sizes in time and space. 

The initial conditions are imposed as a Gaussian wave packet placed to the right of the potential. 
Its initial motion is restricted solely in the direction of the potential ensured by the advection equation
\begin{align}\label{eq:initial_data}
\Phi(0, x)  &= B e^{-\left(\frac{x-\mu}{2d}\right)^2}\,,
\\
\label{eq:initial_data2}
\frac{\text{d} \Phi(0, x)}{\text{d} t} &= \odv{\Phi(0, x)}{x}\,.
\end{align}
We choose outgoing boundary conditions by applying advection equations at the grid's edge points. 
However, the boundaries are positioned sufficiently far from the region of interest so that the choice of boundary conditions has no impact on the results for the length of the analyzed signals. 

After performing convergence tests, we specify the initial data parameters' values to $B=1, \mu=30M$, and $d=1M$. 
We choose the end time $300M$ and the spatial domain between $x=-300M$ and $x=300M$.

\begin{figure}
    \centering
    \includegraphics[width=\linewidth]{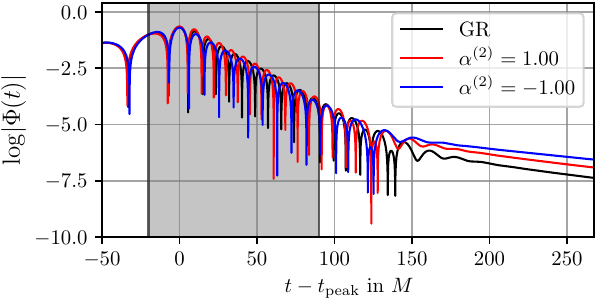}
    \caption{Exemplary waveforms produced from the same initial data given from Eqs.~\eqref{eq:initial_data} and~\eqref{eq:initial_data2} and from different potentials, GR (black), with $\alpha^{(2)}=1.00$ (blue) and with $\alpha^{(2)}=-1.00$ (red). 
    The gray-shaded area indicates a typical portion included in the fits.
    }
    \label{fig:long_waveforms}
\end{figure}

\subsection{Ringdown modeling}\label{ss:rd_modeling}

While there is ongoing discussion regarding the onset of linear perturbation theory, the influence of nonlinearities, and the significance of higher-order perturbations~\cite{Giesler:2019uxc,Sberna:2021eui,Cheung:2022rbm,Mitman:2022qdl,Konoplya:2022pbc,Baibhav:2023clw,Nee:2023osy,Cheung:2023vki,Zhu:2023mzv,Redondo-Yuste:2023ipg,Silva:2024ffz,Giesler:2024hcr,Mitman:2025hgy}, it is well established that at late times, the QNMs --- exponentially decaying sinusoids --- provide a good description of the ringdown. 
Since the decay time of the first overtone is approximately 3 times shorter than that of the fundamental mode and even smaller for higher overtones, the time domain ringdown can be accurately described using only the fundamental mode at sufficiently late times. 
However, it is also well known that the dominant contribution at late times is given by a power-law tail, which is often not included when extracting QNMs at intermediate times. 

The waveform signal that we study is obtained by fixing the radius of the solution, ${\Phi}(t) = \Phi(t, x=R)$. 
For standard comparison with other works, we rescale the time such that the maximum of the waveform is identified as $t_\mathrm{peak}$, we show some examples of related waveforms in Fig.~\ref{fig:long_waveforms}. 
To extract the QNMs, we use a theory-agnostic model describing the signal as
\begin{align}\label{eq:TA_model}
{\Phi}(t) =   &\sum_{n=0}^{N-1} A_n e^{\left(-\omega^\mathrm{im}_n(t-t_\mathrm{peak})\right)}
\sin\left( \omega^\mathrm{re}_n (t-t_\mathrm{peak}) + \phi_n \right)
\nonumber\\ 
\end{align}
This model uses 4 free parameters to model each QNM, namely the amplitude $A_n$, the phase $\phi_n$, and the real part $\omega^\mathrm{re}_n$ and the imaginary part $\omega^\mathrm{im}_n$ of the complex QNM frequency. 
To also model the late-time tail
\begin{align}
\Phi_\mathrm{tail}(t) = \frac{A_\mathrm{tail}}{\left(t-t_\mathrm{tail} \right)^{2\ell+3}}\,,
\end{align}
we need to introduce two more parameters, its amplitude $A_\mathrm{tail}$ and pole $t_\mathrm{tail}$~\cite{Price:1971fb,Leaver:1986gd,Gundlach:1993tp,Barack:1998bw}. 
Thus, in total, we have $4N+2$ free parameters. 
Note that the exponent of the tail depends on the long-range properties of the potential and the type of initial data. 
When introducing nonzero $\alpha^{(2)}$ modifications, the exponent will change. 
However, because we work with perturbative deviations and focus on obtaining the fundamental mode, the modification of the tail exponent will have a small impact on the extracted fundamental mode properties. 
Moreover, to compare the results of all models and modifications, we consider it approximately constant and use the same exponent in all applications. 

To quantify the success of each fitting model in approximating the ringdown waveform, we define the mismatch function between two signals $h_1$ and $h_2$ as
\begin{align}\label{eq:mismtach_def}
    \mathcal{M} = 1 - \frac{\langle h_1, h_2 \rangle }{\sqrt{\langle h_1, h_1 \rangle\langle h_2, h_2 \rangle}}\,,
\end{align}
with 
\begin{align}\label{eq:inner_def}
    \langle h_1, h_2 \rangle = \int_{t_0}^{t_f}h_1(t)h_2(t) \text{d}t\,,
\end{align}
where $t_0$ and $t_f$ define the time window of analysis, with both times measured relative to the signal's peak time.
In Sec.~\ref{applications}, we set $t_f=90M$ and vary $t_0$ from $-20M$ to $50M$. 
The smaller the mismatch, the better the fitting result matches the waveform. 
However, a small mismatch does not automatically describe a good fit. 
If two functions differ by a constant factor, the mismatch does not change. 
Additionally, a small mismatch does not exclude overfitting. 

The procedure to obtain the best-fit parameters is similar to the one outlined in Ref.~\cite{Nee:2023osy}. 
We define a viable range for all fitting parameters to make the numerical optimization more efficient, but it is large enough that our extracted parameters do not exceed those limits. 
To avoid getting stuck in local minima, we repeat the fitting procedure at each starting time using different random initial guesses within the parameter limits. 
These iterations are repeated at least 80 times and are crucial when increasing the number of parameters in the models.

\section{Application and results}\label{applications}

In the following, we show our results for single modifications in Sec.~\ref{app_single} and for multiple modifications in Sec.~\ref{app_multi}. 
The WKB-inspired analysis can be found in Sec.~\ref{app_wkb}.

\subsection{Single modification at a time}\label{app_single}

As a first application, we consider axial perturbations and modifications to the GR potential described by varying one parameter $\alpha^{(k)}$ at a time. 
Note that scalar and polar perturbations in this setup are qualitatively similar and, therefore, not shown here.  
Additionally, the cases with $k\in[4,6]$ are qualitatively similar to those with $k\in[3,5,7]$. 
Therefore, we provide the corresponding results in Appendix~\ref{app_single}.

\subsubsection{Potentials and waveforms}

For a clear presentation, we introduce a single deviation parameter $\epsilon$, which controls the strength of each modification as follows
	\begin{align}
		\alpha^{(k)}(\epsilon) = \epsilon\alpha^{(k)}_\mathrm{max}\,,
	\end{align}
where $\alpha^{(k)}_\mathrm{max}$ is defined from Eq.~\eqref{eq:criterion} as follows
\begin{align}
\alpha^{(k)}_\mathrm{max} = (1+1/k)^k(k+1)\,.
\end{align}
To be within the validity of the perturbative framework, we only consider values of $\epsilon\leq 1$.

\begin{figure*}
    \centering
    \includegraphics[width=\linewidth]{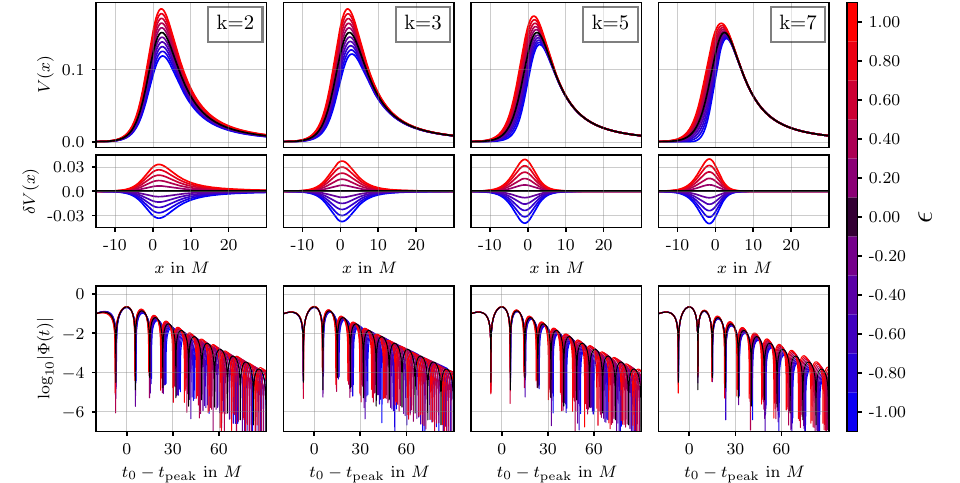}
    \caption{{Modifications to GR potential and respective waveforms.}
    Top row: we show the effective potentials when varying a single $\alpha^{(k)}=\epsilon\alpha^{(k)}_\mathrm{max}$ modification at a time.
   	Middle row: we plot the corresponding deviations from the original GR potential.
    Bottom row: we provide the corresponding waveforms. 
    In each panel, we fix the order $k$ and show a family of curves corresponding to increasing the strength of the modification (different colors). 
    The GR case is shown for comparison (black line). 
    }
    \label{fig:potentials}
\end{figure*}

To get an understanding of how various $\alpha^{(k)}$ modifications deform the effective potential with respect to the GR case, we show them in the top row of Fig.~\ref{fig:potentials}. 
Because of the nature of the  $1/r^k$ modifications, it is evident that the potential is mostly impacted close to the black hole (around the light ring region at about $3M$) and vanishes at the horizon (by construction). 
Increasing the order $k$ shifts the effect of the modifications closer to the black hole horizon and increasingly on the left side of the potential barrier. 
Note that the general structure of a potential with two turning points is not changed for any of the studied cases, e.g., there are no regions with possible bound states and no secondary barriers. 
If those features are present, the validity of the perturbative corrections from the parametrized QNM framework should be questioned.

In the bottom row of the same figure, we also show the waveforms used in our ringdown analysis, which are created by scattering the same initial data with the various modified effective potentials. 
As expected, the waveforms are impacted by the modifications, but there are no nonperturbative features like echoes visible. 
Note that modifications that increase the height of the potential peak cause slightly higher frequencies.

\subsubsection{Mismatches}
 
We now fit the waveforms shown in the lower panels of Fig.~\ref{fig:potentials} with our different ringdown models. 
We vary the starting times and compute mismatches between the time domain signal and the outcome of each fit. 

\begin{figure*}
    \centering
    \includegraphics[width=\linewidth]{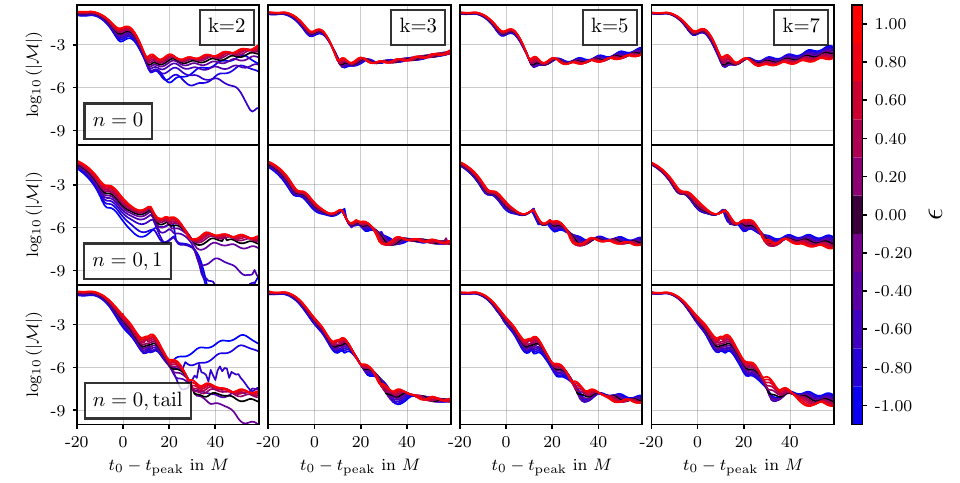}
    \caption{
    {Mismatches between gravitational waveforms and fit results.}
    In the top panels, we consider a model with one QNM; in the central panels, we use two QNMs; and in the bottom panels, we consider one QNM and the power-law tail. 
    From left to right, we consider different powers $k\in[2,3,5,7]$ of the modifications $\alpha^{(k)}$. 
    In each panel, we show mismatches as a function of the starting time $t_0-t_\text{peak}$. 
    Different curves in each panel correspond to different amplitudes of $\alpha^{(k)}$ that are indicated by different colors in the color bar.}
    \label{single_field_mismatches}
\end{figure*}

In each panel of Fig.~\ref{single_field_mismatches}, we analyze ringdown waveforms for a given $k\in[2,3,5,7]$ with numerical values for $\alpha^{(k)}$ indicated in the color bar. 
The qualitative behavior of the mismatches is very similar for $k>2$, but shows more variety for negative values of $k=2$. 

First, we want to discuss the general trend when switching from a single-mode model (top row in Fig.~\ref{single_field_mismatches}) to a model that also includes the first overtone (middle row) or the tail (bottom row). 
The following aspects are universal for $k>2$, but we discuss them separately since individual cases are more complex for $k=2$. 
As one expects, including the overtone reduces mismatches for all start times. 
Qualitatively, it mainly reduces the mismatches at early times but also reduces the plateau at late times. 
Including the Price tail brings modest improvements at early times, but the most significant improvements appear at very late times, as can be seen by yielding the lowest mismatch plateaus. 

Let us make the following observation for $k=2$. 
Here, the larger mismatches for the model, including the fundamental mode and the tail, appear for negative values of $\alpha^{(k)}$, and the opposite trend is visible for the mismatches of the model, including only the fundamental mode and the first overtone. 
This shows that the relative importance of tail and overtone (same initial data) generally also depends on the specific modification of the potential. 
Since the tail depends crucially on the amplitude of the $k=2$ term, increasing/decreasing this contribution consistently impacts our analysis, as reported here. 

\subsubsection{QNM extraction and accuracy}

\begin{figure*}
    \centering
    \includegraphics[width=\linewidth]{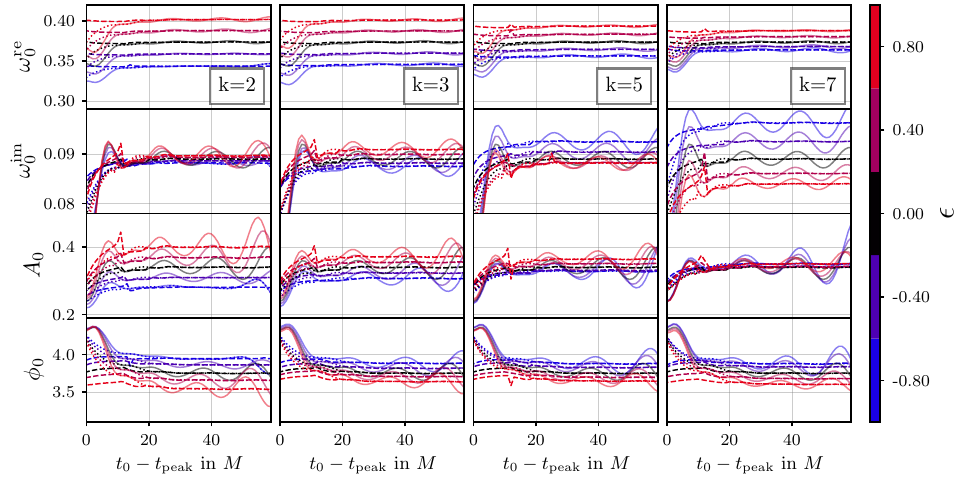}
    \caption{
	{Extracted parameters from gravitational waveforms.}    
    We show from top to bottom the extracted real and imaginary parts of the fundamental QNM, its amplitude $A_0$ and phase $\phi_0$, as function of the starting time $t_0-t_\text{peak}$ when using different fitting models, $n=0$ (solid lines), $n=0,1$ (dashed lines) and $n=0,\mathrm{tail}$ (dotted lines).  
    From left to right, we consider different powers $k\in[2,3,5,7]$ of the modifications $\alpha^{(k)}$. 
}
    \label{fig:params_vs_tshift}
\end{figure*}

Let us turn to one of the central aspects of our study, the extraction of the QNM properties. 
In each row of Fig.~\ref{fig:params_vs_tshift}, we show the extracted values for the four parameters describing the fundamental QNM as a function of the start of the fitting time. 
Different modifications for $\alpha^{(k)}$ are shown across different columns. 
We only show a subset of the analyzed modifications for a cleaner presentation. 
Each panel contains curves in three different line styles corresponding to fits with the $n=0$ model, the $n=0,1$ model, and the $n=0$, tail model. 

In almost all cases, it is evident that the $n=0$ model admits oscillatory behavior around the more stable values. 
The stability of the extracted parameters is significantly improved for the other two fitting models. 
Remarkably, the real part of the fundamental mode is overall in very good agreement with the perturbative predictions and fairly robust for $t_0 - t_\text{peak} \gtrsim 10M$ for all fitting models. 
The imaginary part is much more prone to oscillatory patterns, especially for large $k$. 
The $n=0$ model is quite unreliable overall if evaluated at a single start time, as well as for amplitude and phase. 

\begin{figure}
	\centering
	\includegraphics[width=1.0\linewidth]{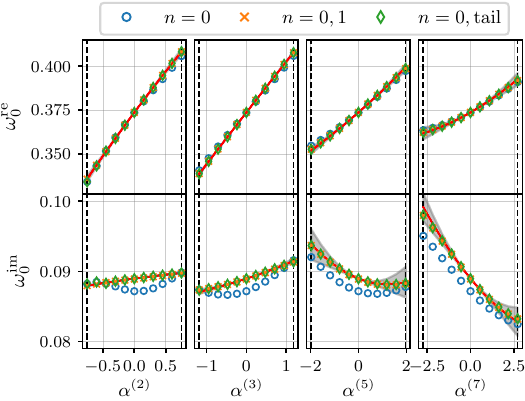}
	\caption{
		{Extraction of the $n=0$ QNM for $t_0 - t_\mathrm{peak}=50M$ and comparison with the perturbative prediction.}
		In the top panels, we show the real part, and in the bottom panels, the imaginary part. 
		From left to right, we consider different powers $k\in[2,3,5,7]$ of the modifications $\alpha^{(k)}=\epsilon \alpha^{(k)}_\mathrm{max}$ and we also indicate each $\alpha^{(k)}_\mathrm{max}$ value (vertical dashed lines). 
		In each panel, we show the perturbative prediction (red line) obtained from Eq.~(\ref{eq:omega_expansion}) and an estimate for its accuracy (grey-shaded region) as a function of the coefficient $\alpha^{(k)}$. 
		Different points correspond to the extracted QNM from the time domain fits of different models. 
	}
	\label{single_field_omega}
\end{figure}

To provide a more careful analysis of the accuracy of the perturbative framework, we show the extraction of the $n=0$ fundamental mode $\omega_{n=0}(\alpha^{(k)})$ in Fig.~\ref{single_field_omega}. 
Here, we analyze the same ringdown waveforms used in Fig.~\ref{single_field_mismatches}. 
In each panel, we show the QNM reconstruction from the time domain and the perturbative prediction at quadratic order from Eq.~\eqref{eq:omega_expansion}. 
Note that the perturbative predictions are not known in cubic order yet, which would be suitable for providing theoretical uncertainties. 
	Instead, we indicate the systematic error coming from the quadratic model by showing the difference between the linear and quadratic model as a gray area, defined between the two curves
\begin{align}
	\omega^\mathrm{err}_{\pm} = \omega \pm \frac{1}{2}\alpha^{(k)}\alpha^{(s)}e_{(ks)}\,,
\end{align}
where $\omega$ is taken from Eq.~\eqref{eq:omega_expansion}. 
As is evident, in all cases using the $n=0,1$ and $n=0,$ tail models, the quadratic prediction is in excellent agreement with the perturbative prediction. 
This strongly underlines the spectral stability of the fundamental mode with respect to the various types of modifications and, thus, a robust connection between time domain and eigenvalues.  
While the real part is even accurate for the $n=0$ model, the imaginary part is not well extracted, not even close to the GR value. 

\subsection{Multiple modifications at a time}\label{app_multi}

In the following, we increase the overall complexity of the problem by considering two cases in which multiple deviation parameters are being varied simultaneously. 
This approach ensures that we also probe the influence of quadratic corrections originating from the couplings of different powers of $k$. 

In our first example, similar to Sec.~\ref{app_single}, we now vary each power as follows 
\begin{align}\label{multi_a_eps}
\alpha^{(k)}(\epsilon) = \epsilon \alpha^{(k)}_\mathrm{max}\,.
\end{align}
where $\alpha^{(k)}_\mathrm{max}$ is defined from Eq.~\eqref{eq:criterion} as follows
\begin{align}
\alpha^{(k)}_\mathrm{max} = (1+1/k)^k(k+1)\,
\end{align}
and we include all the powers $k=2...7$.
To be within the validity of the perturbative framework, we only consider values of $\epsilon$ such that summing over all included powers is sufficiently small. 

Also, in order to change the potential in different ways, we consider a second case where we include only two different powers for $k=2$ and $k=7$ with opposite signs. 
One causes the largest deviation from GR to the real part of the frequency and one causes the largest deviation to the imaginary part, as can be seen in Fig.~\ref{single_field_omega}.
The second modification is defined as
\begin{align}\label{eq:modification_k2277}
	\delta V(r) = \frac{\epsilon}{r_\text{H}^2}\left[\alpha^{(2)}_\text{max} \left(\frac{r_\text{H}}{r}\right)^2 - \alpha^{(7)}_\text{max} \left(\frac{r_\text{H}}{r}\right)^7\right]\,.
\end{align}

\begin{figure}
    \centering
    \includegraphics[width=\linewidth]{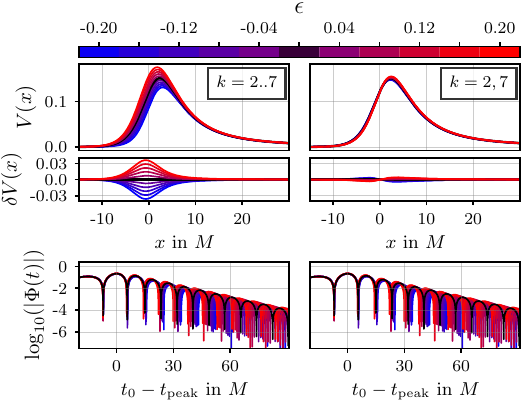}
    \caption{
	{Modifications to GR potential and respective waveforms.}
    Top: we show the effective potential when varying multiple $\alpha^{(k)}$ modifications that include $k=2...7$ (left) and $k=2, 7$ (right), as described in Eqs.~\eqref{multi_a_eps} and~\eqref{eq:modification_k2277} respectively (different colors). 
    The GR case is shown for comparison (black line). 
    Middle: the corresponding deviations from the original GR potential.
    Bottom: the corresponding waveforms produced from the same initial data.
    }
    \label{fig:potentials_multi}
\end{figure}

We start by showing the family of potentials included by these parametrizations in the top panel in Fig.~\ref{fig:potentials_multi}. 
As in the case of single modifications, there is no qualitative change in the two-turning-point nature of the potential barrier, and modifications are mostly around the peak of the barrier and left of it. 
The created waveforms are shown in the bottom panel of Fig.~\ref{fig:potentials_multi}. 
As expected from Sec.~\ref{app_single}, no obvious nonperturbative features exist. 

\begin{figure}
    \centering
    \includegraphics[width=1\linewidth]{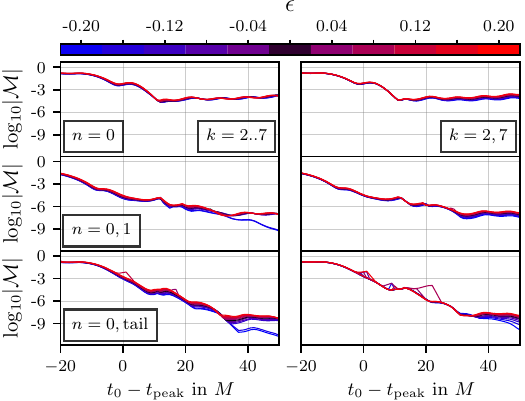}
    \caption{
	{Mismatches between waveforms and fit results.}    
    Here we show similar results as in Fig.~\ref{single_field_mismatches}, but when varying multiple $\alpha^{(k)}$ including $k=2...7$ (left) and $k=2, 7$ (right) at the same time.
    }
    \label{fig:mismatch_many_classic}
\end{figure}

The mismatches are reported in Fig.~\ref{fig:mismatch_many_classic} and qualitatively show the same behavior as for the single modification case. 
Although the $k=2$ term is included, it does not dominate the behavior, as in the case of the single modifications; instead, the overall behavior is similar to that of $k > 2$. 

\begin{figure}
    \centering
    \includegraphics[width=1\linewidth]{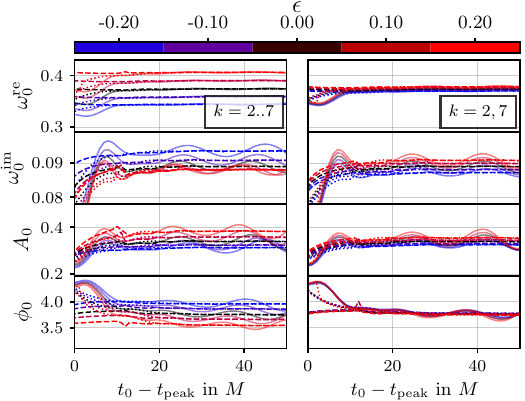}
    \caption{
	{Extracted parameters from gravitational waveforms.}    
    Here we show similar results as in Fig.~\ref{fig:params_vs_tshift}, but when varying multiple $\alpha^{(k)}$ including $k=2...7$ (left) and $k=2, 7$ (right) at the same time.
    }
    \label{fig:params_vs_tshift_many}
\end{figure}

\begin{figure}
    \centering
    \includegraphics[width=1\linewidth]{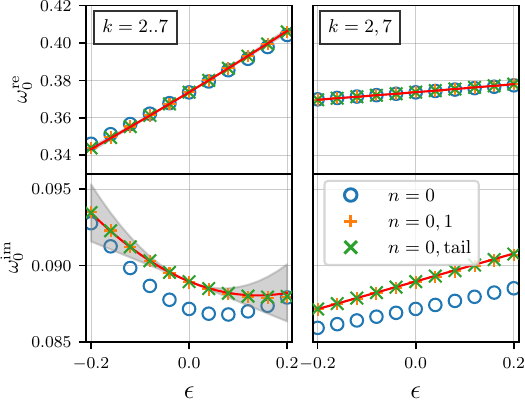}
    \caption{
	{Extraction of the $n=0$ QNM for $t_0-t_\mathrm{peak}=50M$ and comparison with the perturbative prediction.}    
    Here we show similar results as in Fig.~\ref{single_field_omega}, but when varying multiple $\alpha^{(k)}$ including $k=2...7$ (left) and $k=2, 7$ (right) at the same time.}
    \label{fig:qnms_many}
\end{figure}

The extraction of the fundamental QNM parameters as a function of the fit starting time is shown in Fig.~\ref{fig:params_vs_tshift_many}.
Note that the extracted QNM parameters remain very stable at late times when the the model includes either the first overtone or the late-time tail and without these components the parameters show oscillatory behavior.
In Fig.~\ref{fig:qnms_many}, we show the extracted fundamental mode for a given starting time at $t_0-t_\mathrm{peak}=50M$.
The extracted QNM is in excellent agreement with the perturbative prediction and remains well within the uncertainty range when the model includes either the first overtone or the late-time tail. 
However, the $n=0$ model fails to predict the frequencies due to its strong dependence on the fit starting time.
This disagreement of the single mode model can be traced back to the late-time oscillations of the parameters caused by the existence of the tail, which can be observed in Fig.~\ref{fig:params_vs_tshift_many}.


\subsection{WKB-inspired analysis}\label{app_wkb}

In the following, we first provide some context about the WKB method and then explain how it can be used to analyze our time domain waveforms. 

It is well known that black hole QNMs are related to the local properties of the perturbation potential. 
This connection has been established in a variety of works; including approximate potentials like quadratic, or P\"oschl-Teller potentials~\cite{Mashhoon:1982im,BLOME1984231,Ferrari:1984ozr,Ferrari:1984zz,Volkel:2022ewm}, as well as more systematically by using WKB theory~\cite{Schutz:1985km,Iyer:1986np,Iyer:1986nq,Kokkotas:1988fm,Seidel:1989bp,Konoplya:2003ii,Matyjasek:2017psv}, which at leading order yields the Schutz-Will formula
\begin{align}
\label{schutz-will}
\omega_n^2 = V^{(0)} - i \left(n+\frac{1}{2} \right) \sqrt{-2V^{(2)}}\,,
\end{align}
where $V^{(0)}$ is the height of the potential at its peak and $V^{(2)}$ the second derivative with respect to the tortoise coordinate $x$, also evaluated at the peak~\cite{Schutz:1985km}. 
This result is not limited to GR but is valid for any potential barrier with two turning points for which WKB theory is a good approximation. 
Higher orders in WKB theory contain corrections with respect to higher derivatives, which are especially important for overtones. 
A complementary conclusion can be drawn in the eikonal limit ($\ell \rightarrow \infty $) that connects the orbital period and Lyapunov exponent of circular photon orbits to the fundamental mode ~\cite{Press:1971wr,1972ApJ...172L..95G}. 

In the context of the parametrized QNM framework, it was shown that knowing a few QNMs allows one to constrain the local properties of the potential peak~\cite{Volkel:2022khh}. 
This even holds when many deviation parameters are varied simultaneously, in which case the individual parameters cannot be constrained, but only their correlations.  
In the same work, using higher-order WKB theory, the coefficients of the parametrized framework have also been mapped to small deviations of the effective potential in terms of deviations of its derivatives at the peak. We therefore expect that, if the extraction of QNMs from the time domain is accurate, it should exhibit similar properties. 

We now ask the following question: Given $\omega_n$ from a ringdown observation, can one use the Schutz-Will formula~\eqref{schutz-will} for accurately inferring the height of the modified potential $V^{(0)}$ and its second derivative $V^{(2)}$?

A closer look at the real and imaginary part of the Schutz-Will formula~\eqref{schutz-will} (for $n=0$) allows one to find $V^{(0)}$ and $V^{(2)}$ as a function of $\omega^\mathrm{re}$ and $\omega^\mathrm{im}$
\begin{align}
\label{V0_SW}
V^{(0)}  &= {(\omega^{\mathrm{re}})}^2 -  {(\omega^{\mathrm{im}})}^2\,,
\\
\label{V2_SW}
V^{(2)} &= - 8 \left( \omega^\mathrm{re} \omega^\mathrm{im} \right)^2\,.
\end{align} 
Note that, although the prediction for $V^{(0)}$ depends on $\omega^\mathrm{im}$, its contribution in the case of Schwarzschild $\ell=2$ and $n=0$ is about half an order of magnitude smaller than the one from the real part (see e.g. Fig.~\ref{single_field_omega}). 
This suggests that the height of the potential should be rather insensitive to typical modifications of the imaginary part. 
In contrast, the second derivative of the potential should be proportional to the real and imaginary parts.  

If one is interested in small deviations from GR, one can use Eqs.~(\ref{V0_SW}), (\ref{V2_SW}) to predict the linear Taylor coefficient of $V^{(0)}$ as a function of $\omega^\mathrm{re}$ and of $V^{(2)}$ as a function of $\omega^\mathrm{re} \omega^\mathrm{im}$. 
Note that in the following, we are only interested in studying the Taylor expansion of $V^{(0)}$ and $V^{(2)}$ along these specific directions because they are motivated from the structure of the Schutz-Will formula. 
This yields the following prediction
\begin{align}
\label{V0bGR}
V^{(0)}(\omega^\mathrm{re}) &\approx V^{(0)}_\text{GR} + 2 \omega_\mathrm{GR}^\mathrm{re} \delta \omega^\mathrm{re}\,,
\\
\label{V2bGR}
V^{(2)}(\omega^\mathrm{re} \omega^\mathrm{im}) &\approx V^{(2)}_\mathrm{GR} - 16 \omega^\mathrm{re}_\text{GR} \omega^\mathrm{im}_\text{GR} \delta \omega^\mathrm{re,im}\,,
\end{align}
with $\delta \omega^\mathrm{re} = \omega^\mathrm{re}-\omega^\mathrm{re}_\text{GR}$ and $\delta \omega^{\mathrm{re,im}} = \omega^\mathrm{re} \omega^\mathrm{im}-\omega^\mathrm{re}_\text{GR} \omega^\mathrm{im}_\text{GR}$.
Note that the linear correction to $V^{(0)}$ is independent of $\omega^\mathrm{im}$. 
One might also use only the Schutz-Will formula without ``calibrating'' to the GR case, but that would introduce a finite error that Eq.~\eqref{V0bGR} avoids and a dependency on $\omega^\mathrm{im}$. 

\begin{figure}[t]  
    \centering
    \begin{minipage}{\linewidth}
        \centering
        \includegraphics[width=\linewidth]{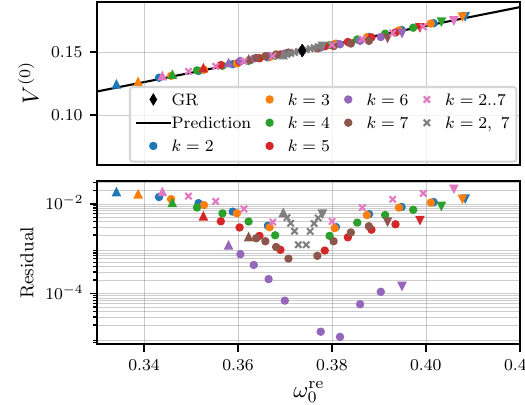}
        \vspace{0.2cm} 
    \end{minipage}
\vskip -1em
    \begin{minipage}{\linewidth}
        \centering
        \includegraphics[width=\linewidth]{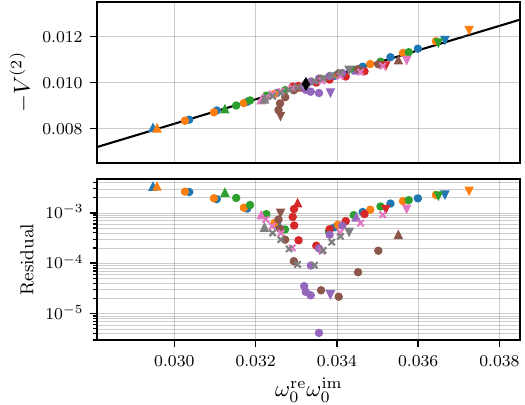}
    \end{minipage}
    \caption{
    {Correlations of potential properties with fundamental mode parameters.}
    We show the correlations of the height $V^{(0)}$ (first panel) and second derivative $V^{(2)}$ (third panel) of the effective potential with respect to the real and imaginary part of the $n=0$ QNM for various $\alpha^{(k)}$. 
    We also show the residuals (second and fourth panel), the absolute difference between the exact values of the potential properties and the ones calculated from Eqs.~(\ref{V0_SW}) and (\ref{V2_SW}) using the extracted frequencies.
    Up and downward triangles indicate the smallest and largest value of $\epsilon$, respectively.
    All data points correspond to those of the single and multiple modifications at a time when using the fitting model with fundamental mode and power-law tail with $t_0-t_\mathrm{peak}=50M$.
    }
    \label{wkb_1}
\end{figure}

In Fig.~\ref{wkb_1}, we show how the properties of the potential correlate with the extracted real and imaginary part of the fundamental mode. 
In the same figure, we also provide residuals for a more quantitative comparison. 
We use results from all the cases studied in Secs.~\ref{app_single} and ~\ref{app_multi}. 
As evident from the top panel, the correlation between the height of the potential $V^{(0)}$ and the extracted fundamental mode is excellent; it is very robust throughout different $k$ and shows a linear behavior. 
The relation between $V^{(2)}$ and the product of real and imaginary parts of the fundamental mode is quite robust for most $k$, but shows a nonlinear behavior for $k\in[6,7]$, indicating that the imaginary part is more sensitive to higher powers of $1/r^{k}$. 

We also show the analytic predictions from Eqs.~\eqref{V0bGR} and~\eqref{V2bGR} for comparison.  
The agreement for $V^{(0)}$ is excellent in all cases. 
For $V^{(2)}$ it is very good for most $\alpha^{(k)}$ modifications, but it becomes less accurate for large $k$, i.e., we find clear deviations for $k\ge6$. 
We expect that higher-order WKB corrections, including higher derivatives of the effective potential, will become more relevant and, therefore, cause inaccuracies.  

In Appendix~\ref{full_correlations}, we show a more comprehensive set of correlations, including properties of the tail, as well as QNM amplitudes. 
While the connection is well established for the QNM spectrum, it has been less studied for the excited amplitudes and phases. 
Since those depend on the specific initial data and the details of the potential through excitation factors and excitation coefficients, they are not purely intrinsic. 
Although we find strong correlations when increasing $\alpha^{(k)}$, they are overall less universal throughout different $k$.

\section{Discussion}\label{discussion}

We discuss the validity of the parametrized QNM framework in the time domain in Sec.~\ref{disc_validity}, the importance of overtones and tails in Sec.~\ref{disc_overtones_tail}, and future directions in Sec.~\ref{disc_future}.

\subsection{Validity of parametrized QNM framework}\label{disc_validity}

Because of ongoing investigations regarding the spectral stability of black hole QNMs, or lack thereof, it is essential to quantify whether theory-agnostic parametrizations have predictive power for ringdown tests. 
As demonstrated in Secs	.~\ref{app_single} and ~\ref{app_multi}, the perturbative predictions for the $n=0$ QNM can be robustly extracted from the time domain, i.e., they are well within the modeling uncertainty of the perturbative predictions. 
This establishes strong evidence that the phenomenological deviations introduced in the parametrized QNM framework can also be used for time domain studies. 

Moreover, our analysis provides independent verification that the numerical coefficients provided in Ref.~\cite{Volkel:2022aca} are accurate and not the limiting factor for ringdown analyses, contrary to the unclear starting time of the fit, the number of included QNMs and the modeling of the power-law tail.

\subsection{Importance of overtone and tail}\label{disc_overtones_tail}

Theory-agnostic ringdown tests are very sensitive to the number of included QNMs, especially overtones.  
By including the power-law tail at late times, which is not done routinely, we have demonstrated that the QNM parameters can be extracted more accurately. 
If not included, the extracted QNM parameters, even for the fundamental mode, show a strong oscillatory dependence on the start of the fitting interval. 
This is also relevant when the power-law tail is not dominating but throughout intermediate times. 
We therefore suggest that power-law tails should be included in high-precision ringdown tests, even if they do not visibly contribute to the analyzed signal. 
Finally, we want to clarify that we are not arguing for using an overtone model at late starting times as a well-motivated physical model. 
Instead, we stress that it improves the accurate extraction of the fundamental mode's properties. 
Systematic errors on QNM parameters due to unmodelled effects have recently been studied in Ref.~\cite{Volkel:2025jdx} and further support our observation.

\subsection{Future directions}\label{disc_future}

One natural extension of this work would be to consider a parametrized QNM framework for rotating black holes beyond GR. 
Recently, such a framework beyond the Teukolsky equation~\cite{Teukolsky:1973ha} has been introduced in Ref.~\cite{Cano:2024jkd}, and the coefficients are known at the linear order. 
However, there are several nontrivial complications that would make a similar extension to the time domain more challenging. 
First, the coupling between the radial and angular equation becomes frequency-dependent itself. 
Second, due to the more complicated structure of the Teukolsky equation, it is less obvious how time dependency could be introduced, and there are fewer theory-specific examples to compare to.

\section{Conclusions}\label{conclusions}

In this work, we introduced a straightforward extension of the parametrized QNM framework~\cite{Cardoso:2019mqo, McManus:2019ulj,Kimura:2020mrh,Volkel:2022aca,Franchini:2022axs,Hirano:2024fgp} to the time domain which allows one to perform spectroscopic studies of phenomenologically modified black holes. 
By considering many possible modifications, we systematically verified that the perturbative corrections to the fundamental QNM can be robustly extracted from a theory-agnostic ringdown fitting procedure; see Figs.~\ref{single_field_omega} and ~\ref{fig:qnms_many}. 
Our analysis demonstrates that the modifications introduced in the parametrized QNM framework are reliably present in the time domain. 
It also shows that the extracted QNMs are in excellent agreement with perturbative predictions in quadratic order. However, this typically requires the introduction of one overtone or the Price tail (leading power-law tail) to the fitting model. 
This result is relevant for ongoing studies investigating the stability of the QNM spectrum itself~\cite{Nollert:1996rf,Nollert:1998ys,Barausse:2014tra,Jaramillo:2020tuu,Jaramillo:2021tmt,Cheung:2021bol,Courty:2023rxk}. 

We also report two important aspects related to the power-law tail~\cite{Price:1971fb,Leaver:1986gd,Gundlach:1993tp,Barack:1998bw}, which becomes dominant at very late times in a ringdown. 
First, including the Price tail, even if it does not dominate the analyzed signal at intermediate times, has an important impact on the accuracy of extracted parameters. 
Not including the power-law tail gives rise to a strong oscillatory behavior of the extracted QNM parameters as a function of the fit starting time.
This behavior is drastically reduced once the tail is included, as can be seen in Figs.~\ref{fig:params_vs_tshift} and ~\ref{fig:params_vs_tshift_many}.

Second, because the power-law tail is strongly related to the long-range decay of the effective potential, we report that modifications, i.e., through the parametrized QNM framework, can have very strong effects on the relative importance of overtones and the tail contribution, even at intermediate times. 
We find that mismatches obtained from models, including the fundamental mode and either one overtone or the power-law tail, can improve by several orders of magnitude, relative to the single-mode model; see Figs.~\ref{single_field_mismatches} and ~\ref{fig:mismatch_many_classic}. 
A detailed analysis of how power-law tails change as a function of modification to the effective potential was studied in Ref.~\cite{Rosato:2025rtr}.
We note that the role of tails in numerical relativity simulations has received much attention recently and may leave observational signatures in some parts of the binary black hole parameters space~\cite{Carullo:2023tff,Cardoso:2024jme,DeAmicis:2024not,Islam:2024vro,Ma:2024bed,DeAmicis:2024eoy,Ma:2024hzq}. 

Finally, inspired by WKB theory, in particular the leading-order result known as Schutz-Will formula~\cite{Schutz:1985km}, we report a semianalytic result for the local properties of the effective potential. 
For small deviations from GR, its height at the maximum can be robustly parametrized with only the real part of the fundamental QNM, while its second derivative with respect to the tortoise coordinate depends strongly on the product of real and imaginary parts; see Fig.~\ref{wkb_1}. 
This demonstrates that ringdown tests based on fitting damped sinusoids in setups beyond GR, like the one studied here, are able to put tight constraints on the local properties of the effective potential. 
This agrees with and supports a related study based on a Bayesian analysis using QNMs directly as input, which bypassed the actual extraction problem~\cite{Volkel:2022khh}. 
We leave the extension of building robust relations using higher-order WKB results for future work.

After our work had been submitted, Ref.~\cite{deMedeiros:2025ayq} appeared. It provides a complementary analysis of the time evolution in similar modified potentials and includes additional spin fields. The QNMs are computed with a continued
fraction method instead of the perturbative parametrized QNM framework.

\acknowledgments
S.~T. and S.~H.~V. thank Nicola Franchini for useful discussions and valuable comments on the manuscript. 
S.~T. acknowledges the Max Planck Institute for Gravitational Physics (Albert Einstein Institute) for hospitality while carrying out parts of this work. 
S.~H.~V. acknowledges funding from the Deutsche Forschungsgemeinschaft (DFG) under Project No. 386119226. 

\section*{DATA AVAILABILITY}
The data that support the findings of this article are not publicly available. The data are available from the authors
upon reasonable request.

\appendix

\section{Additional single modification powers}\label{rest_single_mod_k}

\begin{figure}[h!]
	\centering 
	\includegraphics[width=\linewidth]{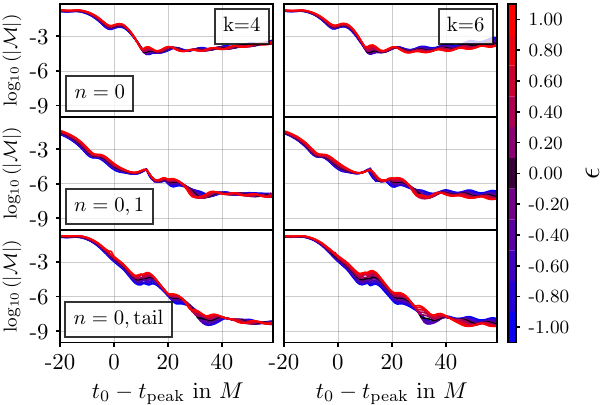}
	\caption{Mismatches as a function of the starting time of the fit, using the three different models, corresponding to single power modifications for $k\in[4,6]$.}
	\label{fig:mismatch_k46}
\end{figure}

\begin{figure}[h!]
	\centering
	\includegraphics[width=\linewidth]{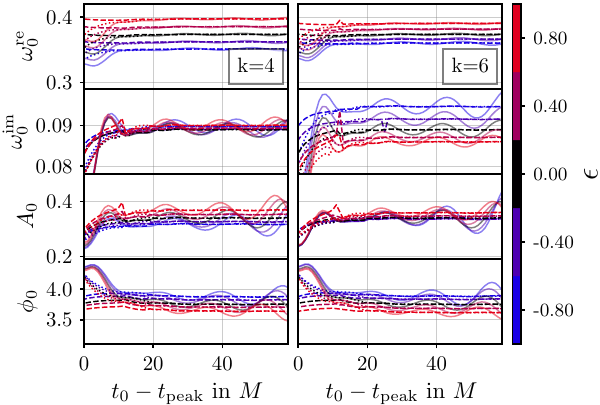}
	\caption{Extracted parameters of the fundamental mode as a function of the starting time of the fit, using the three different models, corresponding to single power modifications for $k\in[4,6]$.}
	\label{fig:params_vs_t0_k46}
\end{figure}

Here, we report the results for the single modification powers $k\in[4,6]$. 
In Fig.~\ref{fig:mismatch_k46}, we show the mismatch as a function of the starting time of the fit for the three different models $n=0$, $n=0, 1$ and $n=0$, tail. 
Additionally, in Fig.~\ref{fig:params_vs_t0_k46}, we show the corresponding results for the extracted parameters.
We can observe that they are qualitatively similar to the ones corresponding to $k\in[3, 5, 7]$, presented in Sec.~\ref{app_single}.

\section{Additional correlations}\label{full_correlations}

\begin{figure*}[]
    \centering
    \includegraphics[width=1.0\linewidth]{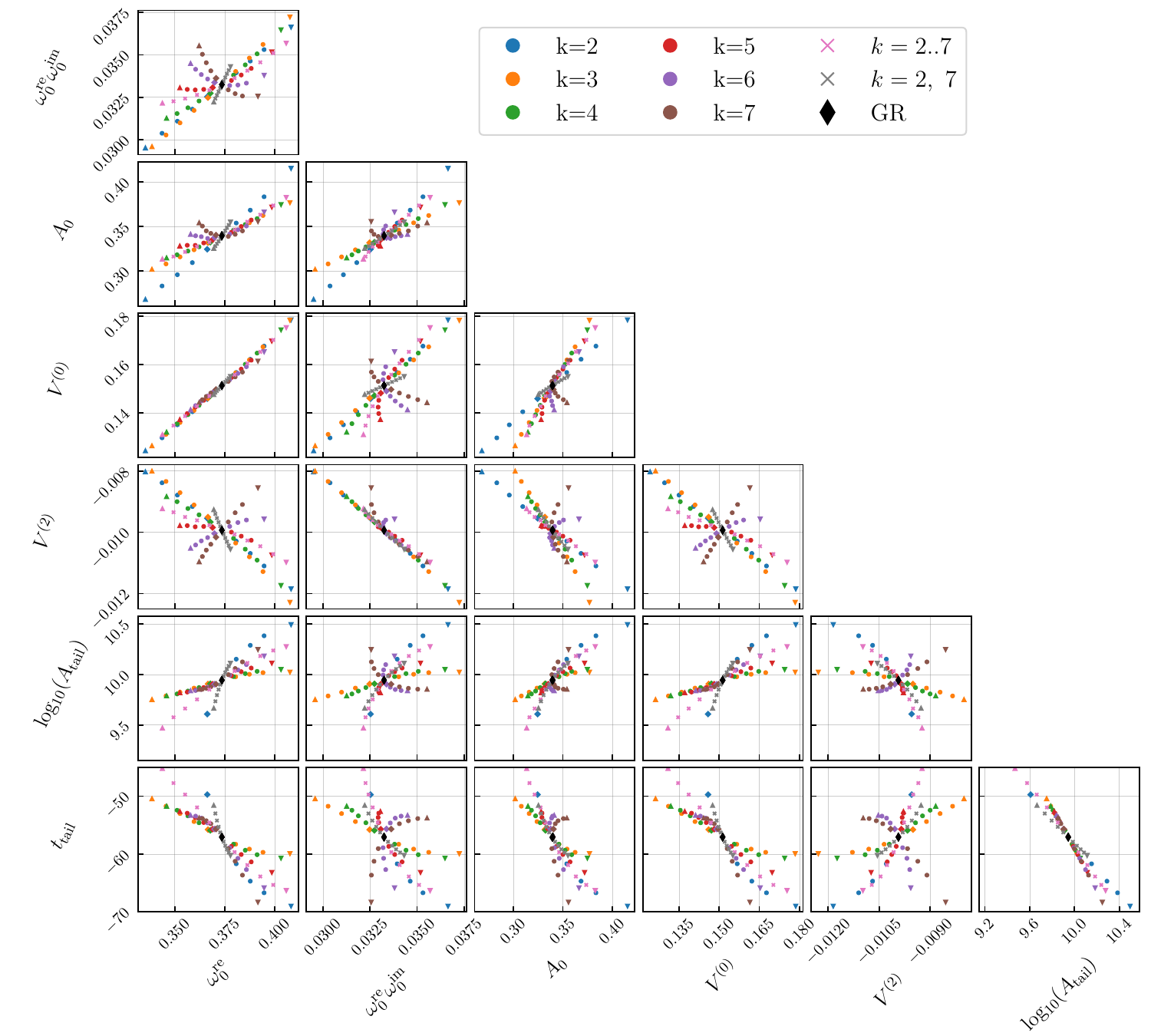}
    \caption{Here, we report a more comprehensive set of correlations between the fitting parameters of the fundamental mode and the power-law tail. 
     Up and downward triangles indicate the smallest and largest value of $\epsilon$ , respectively.
    We use the $n=0,\mathrm{tail}$ model at $t_0-t_\mathrm{peak}=50M$ to obtain the data points. \label{fig_more_correlations}}
\end{figure*}

The following provides a more comprehensive overview of additional correlations between the various QNM models and properties of the potential barrier. 
In Fig.~\ref{fig_more_correlations}, we include the correlations of our fitting results with respect to the amplitude of the fundamental mode $A_0$, the amplitude of the power-law tail $A_\text{tail}$, and its pole $t_\text{tail}$. 
The previously described correlations between $V^{(0)}$ and $\omega^\mathrm{re}_0$ stand out. 
There is also a remarkable correlation between $\log_{10}(A_\mathrm{tail})$ and $t_\mathrm{tail}$. 
Finally, note that $V^{(2)}$ and $t_\mathrm{tail}$ are very sensitive to the power of $k$. 

\newpage

\bibliography{literature}

\end{document}